\documentclass[letterpaper,journal]{IEEEtran}

\usepackage{amsmath,amsfonts} % 只加载一次
\usepackage{textcomp}

\usepackage[ruled,vlined,linesnumbered]{algorithm2e}

\SetKwFor{For}{for}{do}{end}
\SetKwFor{ForEach}{for each}{do}{end}
\SetKwFor{While}{while}{do}{end}
\SetKwInput{KwInput}{Input}
\SetKwInput{KwOutput}{Output}
\SetKw{KwTo}{to}

\usepackage{graphicx}
\usepackage{subcaption} 
\usepackage[font=small,labelfont=bf]{caption}
\usepackage[table]{xcolor}
\usepackage{tikz}
\usetikzlibrary{arrows.meta,positioning,fit,calc,shapes,backgrounds}

\usepackage{dblfloatfix}

\usepackage{array}
\usepackage{tabularx}
\usepackage{booktabs}
\usepackage{multirow}

\usepackage{booktabs}
\usepackage{multirow}
\usepackage{makecell}
\usepackage{graphicx}
\usepackage{graphicx}      % 为 \resizebox
\usepackage{booktabs,multirow,makecell}

\usepackage{url}
\usepackage{verbatim}
\usepackage{cite}

\usepackage[hidelinks]{hyperref}

\definecolor{BestBlue}{HTML}{1F77B4}
\definecolor{WorstRed}{HTML}{D62728}

\newcommand{\bb}[1]{\textbf{\textcolor{BestBlue}{#1}}}

\newcommand{\key}{\ensuremath{k}}

\newcommand{\rr}[1]{\textbf{\textcolor{WorstRed}{#1}}}

\begin{document}

\title{Plaintext Structure Vulnerability: Robust Cipher Identification via a Distributional Randomness Fingerprint Feature Extractor}

\author{%
  Xiwen~Ren, Min~Luo, Cong~Peng, and Debiao~He%
  \thanks{(\textit{Corresponding authors}: Min Luo; Cong Peng.)}%
  \thanks{Xiwen Ren is with the School of Cyber Science and Engineering, Wuhan University, Wuhan 430072, China (e-mail: rxiwen@whu.edu.cn).}%
  \thanks{Min Luo is with the School of Cyber Science and Engineering, Wuhan University, Wuhan 430072, China (e-mail: mluo@whu.edu.cn).}%
  \thanks{Cong Peng is with the School of Cyber Science and Engineering, Wuhan University, Wuhan 430072, China (e-mail: cpeng@whu.edu.cn).}%
  \thanks{Debiao He is with the School of Cyber Science and Engineering, Wuhan University, Wuhan 430072, China, and also with the Shanghai Key Laboratory of Privacy-Preserving Computation, Matrix Elements Technologies, Shanghai 201204, China (e-mail: hedebiao@163.com).}%
}

% \markboth{IEEE TRANSACTIONS ON INFORMATION FORENSICS AND SECURITY}%


\maketitle

\begin{abstract}
Modern encryption algorithms form the foundation of digital security. However, the widespread use of encryption algorithms results in significant challenges for network defenders in identifying which specific algorithms are being employed. More importantly, we find that when the plaintext distribution of test data departs from the training data, the performance of classifiers often declines significantly. This issue exposes the feature extractor's hidden dependency on plaintext features. To reduce this dependency, we adopted a method that does not learn end-to-end ciphertext bytes. Specifically, this method is based on a set of statistical tests to compute the randomness feature of the ciphertext, and then use the frequency distribution pattern of this feature to construct the algorithms' respective fingerprints. The experimental results demonstrate that our method achieves high discriminative performance (e.g., AUC $>$ 0.98) in the Canterbury Corpus dataset, which contains a diverse set of data types. Furthermore, in our cross-domain evaluation, baseline models’ performance degrades significantly when tested on data with a reduced proportion of structured plaintext. In sharp contrast, our method demonstrates high robustness: performance degradation is minimal when transferring between different structured domains, and even on the most challenging purely random dataset, it maintains a high level of ranking ability (AUC $>$ 0.90).
\end{abstract}


%%%%%%%%%%%%%%%%%%%%%%%%%%%%%%%%%%%%%%%%%%%%%%%%%%%%%%%%%%%%%%%%%%%%%%%%%%%%%%%%%%%%%%%%%%%%%%%%%%%%%%
% 第1段：背景引入 - 介绍密码学算法在现代信息安全中的重要性和算法识别的必要性

\section{INTRODUCTION}
\label{sec:intro}
Modern encryption algorithms are crucial to protecting sensitive data, but a key security gap persists: given only ciphertext, it is still a major challenge to identify which specific algorithm produced it~\cite{meijer2021s}. This uncertainty directly affects compliance verification~\cite{nist2019fips3}. For example, defenders must be able to distinguish between abandoned weak algorithms such as DES and insecure implementations of strong algorithms such as the ECB mode of AES. Therefore, understanding how data is encrypted, not just whether it is, has become crucial for tasks including detecting encrypted ransomware~\cite{erwradar2025}, identifying cryptographic API misuse in real systems~\cite{wang2024cryptody}, and ensuring regulatory compliance~\cite{mousavi2025detecting}.

% 第2段：回顾传统及初步的解决方法，指出传统识别方法的局限性
The challenge of traditional encryption algorithm identification is that only binary documents can be accessed, and these methods mainly rely on traditional heuristics. Among them, detection based on signatures is the most direct strategy: it identifies algorithms by scanning known cryptographic constants such as S-boxes and initialization vectors or specific instruction sequences in binary documents~\cite{meijer2021s, ami2022crypto}. However, if a code obfuscation algorithm is not clearly characterized by itself, it can easily bypass this detection method. In order to overcome this limitation, a more advanced method turns to analyzing the structural characteristics, such as path insensitive emulation with similarity analysis~\cite{bincrypto2025} and LLM-guided semantic retrieval~\cite{foc2024}. Although these early methods were successful in specific scenarios, they have the same limitation: over reliance on manually engineered patterns. This dependence makes them sensitive to compiler optimizations and unknown algorithm variants~\cite{meijer2021s, ami2022crypto}.

% 第3段：介绍更先进的研究趋势及其成果
In order to overcome the limitations of traditional methods, research field has shifted to machine learning to automatically discover unique differences in ciphertext. These methods often follow a two stage process: first, extract a high-dimensional feature vector from the ciphertext, and then input the vector into the classifier for identification. This research is mainly divided into two directions: one direction is to optimize the first stage that extracts richer representative statistical characteristics. For example, it exploits the full distribution of scores from randomness tests such as the NIST suite to detect subtle deviations~\cite{li2025generic}. Another route focuses on the optimization of the second stage, designing more powerful classifiers. Recent studies have revealed that neural network models are effective, for example, CNN-based classifiers perform strongly~\cite{hu2025cnn}. Meanwhile, tree-based integrated models and hybrid models also often achieve high accuracy~\cite{yuan2023,yuan2025mlp}. Overall, both directions report strong accuracy.

% 第4段：总结现有研究的共性问题（研究缺口）
However, a core problem that has been overlooked is whether the trained model captures the inherent attributes of the cryptographic algorithm or depends on the statistical law of the training data. To make this concrete, this distinction is crucial to actual deployment, because the distribution of plaintext in real scenarios is not uniform, including both natural languages with obvious structures and high-entropy data such as compressed documents or random keys. If the model mainly learns the specific statistical characteristics of the plaintext rather than the inherent characteristics of the algorithm, its generalization ability cannot be guaranteed in the new or distributed shifted plaintext structure, which makes it difficult to meet the practical application needs.

% 第5段：提出本文的核心解决思路
Therefore, we propose a feature engineering framework. We do not challenge the indistinguishability of ideal ciphers, but aim to alleviate the impact of Plaintext Structure Vulnerability. This framework produces what we refer to as a cryptographic randomness fingerprint by lifting the analysis from raw bytes to a more stable statistical domain. Specifically, this method constructs a new feature vector by analyzing the ciphertext with a comprehensive set of randomness tests and then capturing the overall shape of their results frequency distribution. This method works because the plaintext mainly influences the raw byte patterns, while the encryption algorithm itself determines the overall statistical characteristics. 

% 第6段：技术实现
To complete this framework, we design a feature extraction with multiple stages. First, the pipeline applies a fixed suite of statistical randomness tests to the ciphertext. Then, to make the diverse results comparable, each raw output is calibrated to a common \( [0,1] \) probability scale, resulting in a structured table of scores. Subsequently, we segment this table of scores into windows. For each window, we then extract robust features by summarizing the score distribution's shape using both histograms and statistical moments. Finally, these distributional summaries from all tests are concatenated to form the final high-dimensional fingerprint. This fingerprint provides a rich and stable input for a wide range of standard classifiers.

In sum, the contributions of this paper are as follows:
\begin{itemize}
  \item \textbf{Ciphertext only feature extractor.} We convert standardized randomness tests into a calibrated score space on \( [0,1] \) via each test's null CDF, and encode distributional shape by using fixed $K$-bin histograms plus $S$ summary scalars. The extractor is plug and play across classical learners and light neural models without task specific tuning.
  
  \item \textbf{Plaintext distribution sensitivity.} We systematically demonstrate that state-of-the-art cipher identification models can degrade sharply under shifts in plaintext distributions. This exposes a core dependency on content statistics rather than algorithmic signal.
  
  \item \textbf{Robustness oriented evaluation.} We standardize a leakage aware protocol with grouped cross validation and boundary purging, fit normalization on training folds only, and report Accuracy, Macro F1, Macro AUC, together with a cross-domain generalization gap to quantify robustness under plaintext and length variation.
  
  \item \textbf{Mechanistic explanation by distributional separability.} Using divergence based analysis in the calibrated space (e.g., inter algorithm divergence and intra class dispersion), we explain why low signal to noise conditions yield the observed pattern of high AUC yet reduced Accuracy and F1.
\end{itemize}

%%%%%%%%%%%%%%%%%%%%%%%%%%%%%%%%%%%%%%%%%%%%%%%%%%%%%%%%%%%%%%%%%%%%%%%%%%%%%%%%%%%%%%%%%%%%%%%%%%%%%%
\section{RELATED WORK}
\label{sec:RELATED WORK}

\subsection{Approaches to Cipher Identification}

Research in automated cipher identification has historically progressed along three main avenues, each with distinct approaches to feature representation and analysis.

\paragraph{Statistical Pipelines}
The statistical pipeline generally consists of two stages. Raw ciphertext is  transformed into a set of statistical methods, which are often derived from the outputs of standardized randomness tests. These features are aggregated into a feature vector and input into a conventional classifier. Representative work in this area includes early schemes based on SVM or decision trees~\cite{dileep2006svm, manjula2011dt}, as well as more recent ensemble methods that leverage randomness test statistics as features~\cite{yuan2023, yuan2025mlp}.

\paragraph{End-to-End Deep Learning}
More recent direction employs end-to-end deep learning to bypass manual feature engineering. Models in this category, typically based on CNN or Transformer architectures~\cite{xie2024}, learn discriminative features directly from raw ciphertext bytes. While these approaches have demonstrated strong accuracy within their training domain, studies show they can be sensitive to shifts in plaintext composition and ciphertext length distributions~\cite{dani2024machine}.

\paragraph{Binary Reverse Engineering}
The avenue approaches the problem from a different modality: binary reverse engineering. Instead of analyzing ciphertext, these methods analyze the cryptographic program's binary code itself. These methods aim to detect the presence of cryptographic functions by analyzing implementation artifacts such as characteristic constants or code patterns within program binaries. They employ static or dynamic analysis techniques, and more recently, large language model–based tools, to automatically identify these cryptographic signatures~\cite{meijer2021s, li2022genda,foc2024}. While targeting a different input, these methods reflect the broader trend towards data-driven identification in this field. Relatedly, another line of code-centric analysis focuses on cryptographic misuse detection, which uses static or dynamic analysis to find implementation-level vulnerabilities, such as insecure cryptographic keys~\cite{li2018khunt,ami2022crypto}. Both code-based approaches underscore the need for ciphertext-only methods when source code is unavailable.

\subsection{Randomness Tests From Verification to Representation}

Randomness test suites such as NIST SP 800-22, Dieharder and TestU01 were originally designed as verification tools for random number quality, rather than components for feature extraction or model training. They calculate the p-values by running a series of statistical tests on the input bit sequence, and give a binary conclusion that whether the sequence is statistically indistinguishable from the ideal random source. These tools emphasize compliance verification~\cite{rukhin2010niststs, brown2006dieharder, lecuyer2007testu01}.

However, a key insight emerged in subsequent research: the full distribution of test statistics or p-values often contains more information than a single verdict. Studies observed that these distributions can exhibit weak yet stable, algorithm-specific patterns, especially under varied data lengths or sampling. This led to a paradigm shift from using tests for verification to using their outputs for representation. Specifically, the modern approach, which we adopt, is to treat the calibrated test scores themselves as features. By mapping all raw, heterogeneous statistics to a common $[0,1]$ probability scale, this method enables richer, distribution-level analysis and across-test comparisons~\cite{vuursteen2023optimal}.

Recent study further supports the strategy of using test outputs as rich features. First, modern combination theory demonstrates that working with a test suite of calibrated scores (like p-values) allows for robust statistical conclusions, even when the tests are not independent~\cite{liu2020cct}. This strengthens the case for using our entire test suite as a unified feature source. Second, research in multi-scale feature design shows that using multiple descriptors to summarize a distribution's shape improves feature stability and separability, which is consistent with our use of both $K$ histogram bins and $S$ summary scalars. Finally, recent work on combining test statistics quantifies the information loss that occurs when compressing many test results into a single number~\cite{vuursteen2023optimal}. This provides a strong motivation for our choice to retain rich, distribution-level summaries rather than collapsing them into a single meta-statistic.

\subsection{Distributional Encoding and Evaluation Protocols}

A recurring design choice in prior pipelines is how to turn the outputs of test suites into features. Many works reduce NIST SP 800-22, Dieharder, or TestU01 to pass or fail like summaries or a handful of single statistics~\cite{rukhin2010niststs,brown2006dieharder,lecuyer2007testu01}, whereas others show that using the \emph{distribution} of calibrated scores or test values carries additional discriminative signal. At the evaluation level, most studies report conventional splits and standardization, with varying degrees of care to avoid refitting on test folds; leakage risks can arise when windowed samples from the same source overlap across folds. On evaluation protocols, leakage-aware splits and OOD-style assessments are encouraged to avoid refitting artifacts and inflated scores; this echoes broader best practices for distribution-shift evaluation~\cite{koh2021wilds}.

\subsection{Research Gap and Positioning}

% 总结文献回顾中发现的研究空白
Literature review shows that there are still the following key gaps in the existing encryption algorithm identification methods:

\begin{enumerate}
    \item \textbf{Methodological Limitation}: Many existing identification methods need to see the inside of the algorithm, mainly relying on binary clues to realize the details. In the black box scenario where only the output can be seen, these prerequisites are not established, and the methods are difficult to play a role.

    \item \textbf{Underutilization of Statistical Information}: Traditional randomness testing mostly adopt the binary judgment of pass/fail, ignoring the rich continuous statistical information that can be used as discrimination characteristics.

    \item \textbf{Limited Pattern Recognition Integration}: 
    Although statistical pattern recognition has proven its application value in some cryptographic problems, the relevant systematic research and its application potential have not been fully explored for the specific scenario of relying on algorithm output for identification.
\end{enumerate}

% 定位本研究对现有研究空白的填补

Our work aims to fill these research gaps: we change the randomness test from a verification tool to a distinctive feature used to depict the characteristics of the algorithm. By analyzing the continuous distribution form of test statistics, instead of simple pass/fail binary results, we can capture the nuances that distinguish between different cryptographic algorithms. The method we proposed uses these patterns, combined with feature engineering and machine learning techniques, to create a method that can be used for algorithmic identification only by output.

%——————————————————————————————————————————————————————————————————————————

\section{PROBLEM DESCRIPTION}
\label{sec:problem_description}
In this section, we formally define the threat model, specifying the ability of adversary and the target of the detector. We also explain the core challenges brought about by the variability of the plaintext structure and define the problem of identifying robust cipher algorithms.

\subsection{Threat Model}
\label{subsec:threat_model}
In our threat model, we consider a scenario where a detector tries to identify the encryption algorithm $A \in \mathcal{A}$ from the observed ciphertext $C$. We assume the ciphertext-only setting: the detector cannot access the plaintext $P$ or the key \(\key\), and the adversary is simply the entity whose encrypted communication is being analyzed.

\paragraph{Adversary's Goal and Nature} The adversary's primary goal is not to attack the detector, but to communicate securely. The adversarial nature is a consequence of their actions, not necessarily their intent. Their choice of plaintext can exploit the Plaintext Structure Vulnerability of learning-based models, causing a detector that is well trained to fail catastrophically on unseen data structures.

\paragraph{Adversary's Capabilities} The adversary's key capability is their full control over the plaintext P. This enables them to generate ciphertext from a diverse range of data sources, thereby creating a significant domain shift problem for the detector:
\begin{itemize}
    \item \textbf{Structured and Low Entropy Plaintext:} Prior learning-based methods have operated on an implicit assumption regarding their input data. They typically use plaintext sources with homogeneous and fixed statistical distributions (e.g., structured text files), a setting which can artificially amplify the subtle statistical artifacts of a given cipher, making identification seem easier than it is in practice.
    \item \textbf{Unstructured and High Entropy Plaintext:} Examples include compressed files, random session keys, or previously encrypted data. This resemblance to true randomness creates a low signal to noise environment that can mask the algorithm’s subtle signature.
\end{itemize}

\paragraph{Design Goals of Detector} To be effective in this setting, the feature extractor $\Phi$ should meet three requirements: (i) invariance to plaintext structure, (ii) sufficiency for distinguishing algorithms, and (iii) stability under common variations such as ciphertext length.

\subsection{Problem Definition}
In this paper, we develop a robust ciphertext-only feature extractor for cryptographic algorithm identification under varying plaintext structures. For clarity, Table~\ref{tab:notation} summarizes the notation used throughout the paper.

Let $C = E_A(P, \key)$ denote the ciphertext produced by algorithm $A$ on plaintext $P$ with key $k$.
Given $C$, the feature extractor $\Phi$ produces a vector $\mathbf{x}=\Phi(C)$; then the classifier $f$ predicts the algorithm label $\hat{y}=f(\mathbf{x})$.

However, the feature vector $\mathbf{x}$ mixes two sources of information: a stable, algorithm-specific part $\mathbf{x}_a$ and a volatile, plaintext-induced part $\mathbf{x}_c$. Therefore, a robust identifier should base its decisions on $\mathbf{x}_a$ rather than $\mathbf{x}_c$.

Accordingly, our goal is to design a feature extractor $\Phi$ whose output is as insensitive as possible to the plaintext distribution $P$.
Formally, given a family of plausible plaintext distributions $\mathcal{P}$, we learn the classifier parameters $\Theta$ and design the extractor $\Phi$ to minimize the worst-case (robust) risk:
\[
\min_{\Theta}\;\sup_{P\in\mathcal{P}}\;\mathbb{E}\big[\ell\big(f_{\Theta}(\Phi(E_A(P,\key))),\,A\big)\big].
\]
Here, $\ell(\cdot,\cdot)$ denotes a standard multi-class classification loss; we use cross-entropy by default.
In practice, we approximate this objective by (i) designing $\Phi$ to be structurally resilient to plaintext shifts (see Section~\ref{sec:methodology}), and by (ii) evaluating under carefully constructed cross-domain splits that vary plaintext composition.

\begin{table}[t]
\centering
\caption{Notation summary}
\label{tab:notation}
\begin{tabular}{l l}
\toprule
Symbol & Meaning \\
\midrule
$\mathcal{A}$ & Set of candidate algorithms \\
$A \in \mathcal{A}$ & True algorithm label \\
$P$ & Plaintext \\
$k$ & Secret key (cryptographic key) \\
$\mathcal{K}$ & Key space \\
$C = E_A(P,\key)$ & Ciphertext produced by $A$ \\
$\Phi$ & Feature extractor (this work) \\
$\mathbf{x}=\Phi(C)$ & Feature vector (fingerprint) \\
$f$ & Classifier mapping $\mathbf{x}$ to labels \\
$W, s$ & Window size and stride \\
$T$ & Number of tests in the suite \\
$K$ & Number of histogram bins \\
$S$ & Number of summary scalars (moments) \\
\bottomrule
\end{tabular}
\end{table}

%%%%%%%%%%%%%%%%%%%%%%%%%%%%%%%%%%%%%%%%%%%%%%%%%%%%%%%%%%%%%%%%%%%%%%%%%%%%%%%%%%%%%%%%%%%%%%%%%%%%%
%方法
\section{METHODOLOGY}
\label{sec:methodology}

\begin{figure*}[t]
    \centering
    \includegraphics[width=\textwidth]{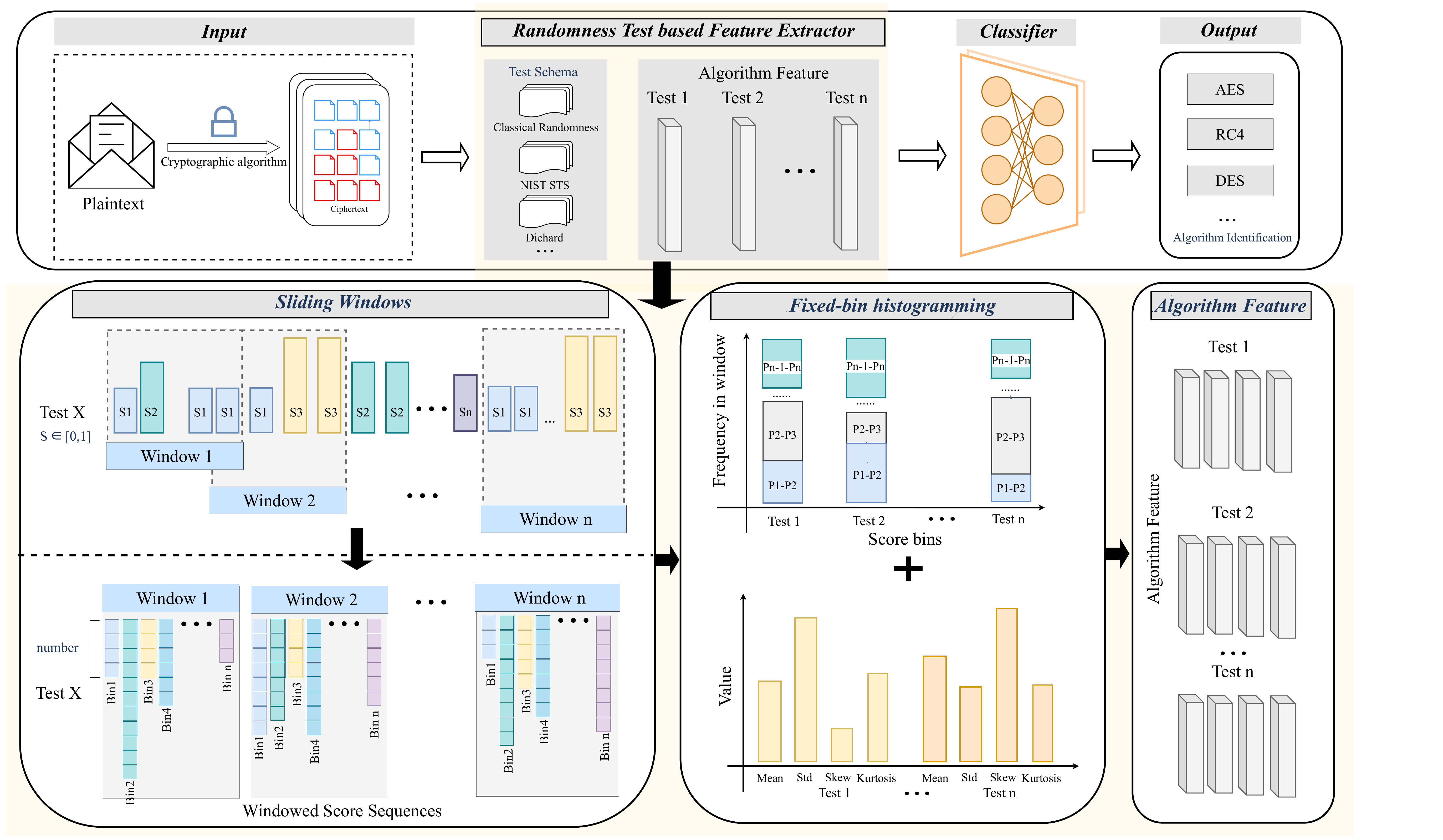}
    \caption{Overview of our test suite distributional feature extractor}
    \label{fig:overview}
\end{figure*}

\subsection{Framework Overview}
\label{subsec:framework}
In this work, we propose a feature extractor that identifies cryptographic algorithms from the statistical distribution of ciphertext outputs, rather than from raw bytes. The core idea is that each cryptographic algorithm leaves a statistical fingerprint that is distinct and stable in the ciphertext. Specifically, our extractor captures these subtle fingerprints while reducing the influence of plaintext structure. As illustrated in Fig.~\ref{fig:overview}, this pipeline consists of four main stages:

\subsubsection{Defining the Feature Columns}
At the first layer, we use a standardized randomness test suite: a fixed panel of tests independent of any specific cipher.
To ensure comparability, the suite has two properties: universality and being fixed in advance, so that the same standard is applied across all datasets and folds.

\subsubsection{Score Calibration}
Next, we map each test’s raw statistic to a calibrated [0,1] score via its theoretical null distribution (CDF).
Consequently, this calibration places heterogeneous test outputs on the same [0,1] scale, enabling across-test comparison and stable downstream aggregation.

\subsubsection{Segmentation and Distributional Aggregation}
Then, we segment the calibrated score table into short windows to capture local patterns, and aggregate each window with fixed-bin histograms and summary moments.

\subsubsection{Fingerprint Assembly and Evaluation}
Finally, we concatenate all window level distributional features in a fixed order to produce a single high dimensional feature vector, which we call a fingerprint.
This fingerprint serves as a standard tabular input for downstream classifiers and is detailed in Sec.~\ref{subsec:fingerprint}.
%————————————————————————————————————————————————————————————————————————————
\subsection{Defining the Feature Columns}
Analyzing raw ciphertext bytes directly is challenging because ciphertext is designed to look random and unstructured. To overcome this, we analyze the ciphertext with a standardized suite of statistical examinations. In other words, rather than working with raw bytes, we operate on the structured and informative outputs of these tests.

Specifically, we implement this strategy by defining a test suite: a diverse collection of $T$ randomness tests that serves as a universal diagnostic toolkit. In our experiments, we instantiate $T{=}41$ tests spanning four complementary families designed to detect different kinds of statistical anomalies:
\begin{itemize}
    \item \textit{Basic statistics} (e.g., mean, entropy) to capture first-order imbalances and tail behavior.
    \item \textit{Classical randomness tests} (e.g., $\chi^2$, runs test) to detect distributional mismatches and run-length irregularities.
    \item \textit{A subset of NIST STS} (e.g., FFT, Linear Complexity) to probe periodicity, compressibility, and template frequencies.
    \item \textit{Advanced suites} (e.g., Diehard’s Birthday Spacings, TestU01’s Serial Correlation) to uncover high-order and long-range dependencies.
\end{itemize}

Having established the composition of the suite, we now formalize its operation. Each test $t \in \mathcal{B}$ maps an input byte sequence $u$ to one or more raw statistics $\{T_{t,j}(u)\}_{j=1}^{m_t}$ at a chosen evaluation length. Evaluating all $T$ tests on an input yields a table of raw statistics with one or more columns per test. This table defines the schema of our first-layer features and becomes the input to the subsequent calibration and aggregation steps.

%————————————————————————————————————————————————————————————————————————————

\subsection{Score Calibration}

To analyze the outputs jointly, we first address a basic issue: the raw statistics lie on heterogeneous scales and are therefore not directly comparable.
We map each raw statistic to a common $[0,1]$ scale via its theoretical null distribution (CDF), using the probability integral transform (PIT).
Formally, for each test $t$ and input sequence $u$, we set
\begin{align}
    s_t(u) &= F_t\!\big(T_t(u)\big) \in [0,1], \label{eq:calibration-def}\\
    s_t(u) &\sim \mathrm{Uniform}(0,1) \quad \text{under } H_0. \label{eq:calibration-dist}
\end{align}
This calibration is essential because it turns heterogeneous raw statistics into standardized scores that behave like p-values.
Consequently, for each input sequence $u$, we collect the calibrated scores into a vector
$\mathbf{s}(u) = [\,s_1(u),\ldots,s_T(u)\,]^\top \in [0,1]^T$, where each entry corresponds to one test.
Then, stacking these vectors across all sequences produces the first layer score table $\mathbf{S}$.
Finally, for numerical stability and reproducibility, we clip each score to $[10^{-12},\,1-10^{-12}]$ and keep a fixed column order across all datasets and folds.

%————————————————————————————————————————————————————————————————————————————

\subsection{Segmentation and Distributional Aggregation}
Next, we segment the calibrated-score table into short windows of length $W$ with stride $s$.
Within each window, and for each test column $t$, we summarize the score distribution by (i) a fixed-$K$ normalized histogram and (ii) $S$ summary moments.
For each test column $t$, we collect the calibrated scores within window $w$ into a vector
$\mathbf{v}_{t,w} \!=\! \big(v^{(1)}_{t,w},\dots,v^{(m_{t,w})}_{t,w}\big) \in [0,1]^{m_{t,w}}$.

Subsequently, we export per-window features by computing a fixed-$K$ normalized histogram and $S$ summary moments for each test column and then concatenating them into a single row-wise fingerprint.

\paragraph*{(i) Fixed-$K$ normalized histogram.}
First, we compute a fixed-$K$ normalized histogram $h_{t,w}$ over $\mathbf{v}_{t,w}$.
Formally, for each bin $b=0,\dots,K-1$, the histogram entry is
\begin{equation}
h_{t,w}[b] \;=\; \frac{1}{m_{t,w}} \sum_{j=1}^{m_{t,w}} \mathbf{1}\!\big\{E_b \le v^{(j)}_{t,w} < E_{b+1}\big\},
\label{eq:hist}
\end{equation}
where $\{E_b\}_{b=0}^{K}$ are the fixed bin edges.
By construction, the histogram entries are normalized so that $\sum_{b=0}^{K-1} h_{t,w}[b]=1$.
A crucial design choice is to fix the bin edges \emph{a priori} (rather than fitting them to data), which helps prevent label or domain leakage into the feature definition.

\paragraph*{(ii) Within-window summary scalars.}
Second, to complement the histogram, we compute a set of $S$ summary scalars that capture the distribution’s key properties.
While the set is flexible—for example, quantiles could be used—our default uses $S\!=\!4$ classical moments (mean, variance, skewness, kurtosis). Specifically,
\begin{subequations}\label{eq:moments}
\begin{align}
    \mu_{t,w} &= \frac{1}{m_{t,w}} \sum_{j=1}^{m_{t,w}} v^{(j)}_{t,w}, \label{eq:moments-mean}\\
    \sigma_{t,w}^{2} &= \frac{1}{m_{t,w}} \sum_{j=1}^{m_{t,w}}
    \bigl(v^{(j)}_{t,w}-\mu_{t,w}\bigr)^{2}, \label{eq:moments-var}\\
    \gamma_{1,t,w} &= \frac{1}{m_{t,w}\,\sigma_{t,w}^{3}} \sum_{j=1}^{m_{t,w}}
    \bigl(v^{(j)}_{t,w}-\mu_{t,w}\bigr)^{3}, \label{eq:moments-skew}\\
    \kappa_{t,w} &= \frac{1}{m_{t,w}\,\sigma_{t,w}^{4}} \sum_{j=1}^{m_{t,w}}
    \bigl(v^{(j)}_{t,w}-\mu_{t,w}\bigr)^{4} - 3. \label{eq:moments-kurt}
\end{align}
\end{subequations}
Here, $\mu_{t,w}$ is the mean, $\sigma_{t,w}^{2}$ the population variance, $\gamma_{1,t,w}$ the skewness, and $\kappa_{t,w}$ the excess kurtosis (Gaussian baseline $=0$).

Taken together, the fixed-bin histogram and the four summary moments provide a \emph{compact, complementary} summary of the score distribution.
\textit{In particular}, the fixed-bin histogram captures the distribution’s global shape by quantifying deviations from the uniform baseline implied by the null hypothesis ($H_0$).
\textit{Meanwhile}, the summary moments encode central tendency, dispersion, asymmetry, and tail weight, and thus remain informative even when some histogram bins are sparse or empty.

%————————————————————————————————————————————————————————————————————————————

\subsection{Fingerprint Assembly and Usage}
\label{subsec:fingerprint}

Finally, we assemble each window’s distributional summaries into a single high dimensional feature vector. Concretely, for every window we concatenate, in a fixed and predefined order, the per test histogram and moment summaries from all $T$ tests. The fixed ordering preserves a stable feature schema across datasets and folds and avoids data dependent changes.

\begin{equation}
\mathbf{x}_w \in \mathbb{R}^{\,T(K+S)} .
\label{eq:fingerprint-dim}
\end{equation}

Instead of doing early feature selection or dimensionality reduction, we leave the discovery of discriminative structure to downstream classifiers. In practice, the fingerprint is used as a standard tabular input for classical learners and light neural models; unless otherwise stated, features are z-score normalized with parameters fitted on the training data only.

%%%%%%%%%%%%%%%%%%%%%%%%%%%%%%%%%%%%%%%%%%%%%%%%%%%%%%%%%%%%%%%%%%%%%%%%%%%%%%%%%%%%%%%%%%%%%%%%%%%%%%%%%%%%%%%%%%%%%%%%%%%%%%%%%%%%%%%%%%%%%%%%%%%%
\section{EXPERIMENTS}
\label{sec:EXPERIMENTS}

\subsection{Experiment Setup}
\label{subsec:exp_setup}

\paragraph{Real-world evaluation on the public Canterbury corpus}
We build a public testbed from the Canterbury Corpus, a diverse collection used in compression research~\cite{canterbury}. It includes English text, source code, HTML, spreadsheets and executables. We split each file into windows of fixed length with a fixed stride. Ciphertexts are generated with standard libraries using per window keys derived by HKDF (RFC 5869) from a master seed and the window identifier~\cite{rfc5869}. This keeps the statistics of real files and makes the algorithm labels reproducible. We release scripts and a manifest to rebuild the ciphertext set from the public corpus.

\paragraph{Synthetic suites with graded structure}
We use five synthetic datasets with graded plaintext structure. They are \texttt{Regular\_100}, \texttt{Regular\_75}, \texttt{Regular\_50}, \texttt{Regular\_25}, and \texttt{Random\_100}. \texttt{Regular\_100} has 100\% regular content.
\texttt{Random\_100} has 0\% regular content. The three mixed sets use 75/25, 50/50, and 25/75 regular to random ratios.
Each dataset contains 10{,}000 windows of 8\,KB. Classes are balanced across six ciphers. All other factors stay fixed, including window size, stride, key policy, and the feature schema. This design isolates the effect of structure and complements the Canterbury evaluation.

\paragraph{Cipher suite and keying policy}
We evaluate six representative ciphers: AES(ECB/CBC), 3DES, Blowfish, ChaCha20, and RC4, covering both block and stream designs, from legacy to modern primitives. Keys and IVs/nonces are per\textendash window and derived, avoiding reuse while ensuring reproducibility. All ciphertext windows have the same fixed length as their plaintext windows.

\paragraph{Baseline}
We benchmark representative linear, tree\textendash ensemble, and neural models on the same fingerprint representation: SVM (linear/RBF), Logistic Regression, Random Forest, XGBoost, MLP, a 1D CNN, a 1D ResNet, and a lightweight Transformer.
Hyperparameters are dataset\textendash agnostic and fixed; standardization is fit on training folds only.

\paragraph{Metrics}
\label{subsec:metrics}
We report Accuracy, Macro F1 score, and Macro AUC. To quantify robustness under distribution shift, we report the cross domain generalization gap, following common practice in distribution shift benchmarks~\cite{koh2021wilds}: for any metric $M \in \{\mathrm{Acc}, \mathrm{F1}_{\text{macro}}, \mathrm{AUC}_{\text{macro}}\}$,
\begin{equation}\label{eq:gap}
\mathrm{Gap}_{s\rightarrow t}(M)
= \frac{M_{s\rightarrow s}-M_{s\rightarrow t}}{M_{s\rightarrow s}} \times 100\%,
\end{equation}
where $M_{s\rightarrow s}$ is the in domain score (train on $s$, test on $s$) and $M_{s\rightarrow t}$ is the cross domain score (train on $s$, test on $t$).
Lower values indicate better cross domain robustness.
All metrics are reported in percentage.

\paragraph{Leakage aware protocol.}
We use a 5$\times$5 stratified group cross validation. Specifically, we group windows by their source file (sample ID) to ensure that all windows from the same file stay in one fold. Additionally, to prevent overlap leakage, we remove a training buffer of $\rho=\lceil W/s\rceil-1$ windows around each test window. Furthermore, we do not use filename or path features, and we fit all data-dependent preprocessing only on the training folds. Finally, we keep the test schema and the global histogram bin edges fixed across all datasets and folds. Three sanity checks confirm this setup: (i) shuffling labels in the training folds drops accuracy to chance level, (ii) fixed bin edges match the performance of refitting edges per fold, and (iii) leaving one cipher out during training still yields high test accuracy.

\subsection{Key Observation of Plaintext Structure}
\label{subsec:Plaintext Structure}
In this subsection, we analyze how plaintext structure drives identification errors from two angles: empirical evidence and statistical characterization.

\begin{figure*}[t]
  \centering
  \includegraphics[width=\textwidth]{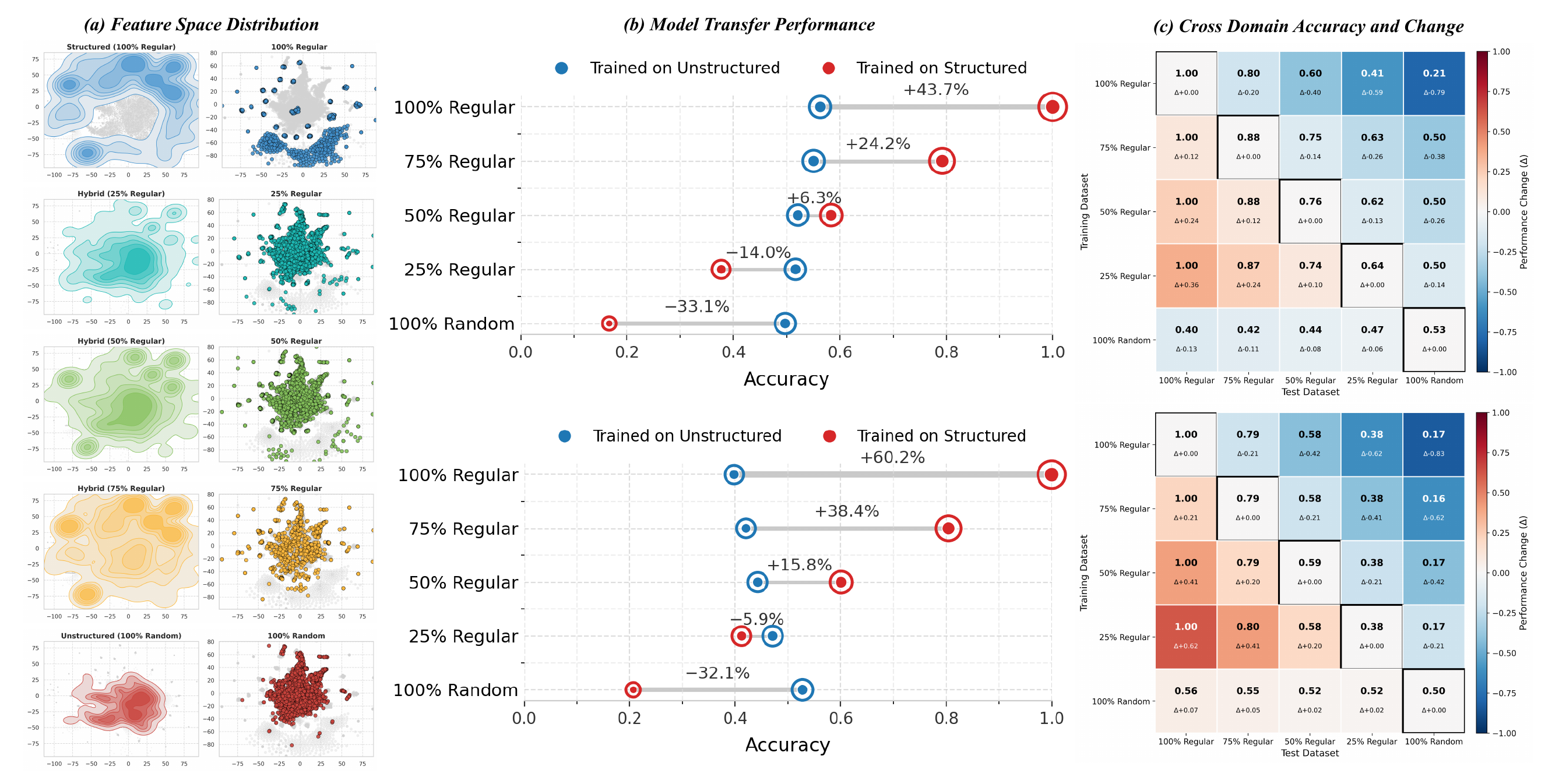}
  \caption{Plaintext structure drives domain gaps.
  (a) t-SNE of fingerprints shows distinct clusters by plaintext regime.
  (b) Cross-domain transfer results reveal sharp drops across regimes.
  (c) An accuracy heatmap summarizes train$\rightarrow$test performance across domains.}
  \label{fig:empirical_evidence}
\end{figure*}
% ——————————————————————————————————————————————————————————————————————————————————————————
\subsubsection{Empirical Validation}
Figure~\ref{fig:empirical_evidence} demonstrates this dependency at a glance.
In panel (a), a t-SNE plot shows that different plaintext regimes form well-separated fingerprint clusters.
In panels (b) and (c), this separation fails to transfer: models that score highly in-domain collapse when evaluated on a different plaintext regime.
These failures indicate that the models have latched onto plaintext specific features instead of algorithm-invariant signals.
% ——————————————————————————————————————————————————————————————————————————————————————————
\subsubsection{Statistical Characterization}
\begin{figure}[t!]
  \centering
  \includegraphics[width=\columnwidth]{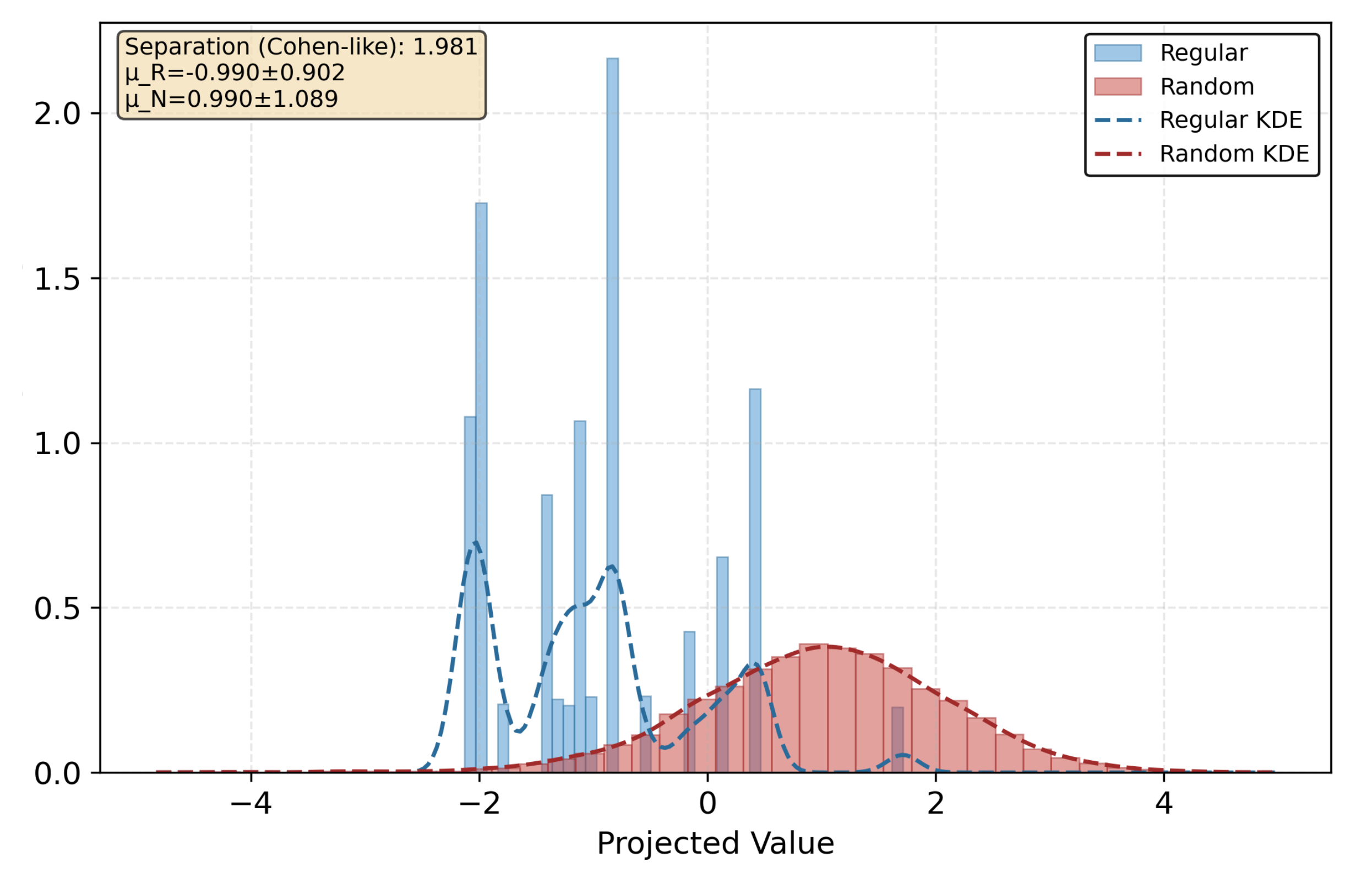}
  \caption{One dimensional LDA projection of the calibrated feature space.
  Despite strong class separation in the full fingerprint space (Bhattacharyya distance $=12.565$),
  the best 1D linear projection achieves only $\approx 1.98$ of separation with substantial overlap,
  indicating that a single axis view cannot preserve the discriminative structure.}
  \label{fig:lda_1d}
\end{figure}

Statistically, the picture is mixed: in the full calibrated feature space the classes separate well, yet along the best linear axis they overlap, which drives the challenge. Concretely, the Bhattacharyya distance in the full space is large (12.565), but after the optimal 1D LDA projection the separation shrinks to $\approx$1.98 with pronounced overlap (see Fig.~\ref{fig:lda_1d}). This high-vs-low dimensional contrast explains the observations: baselines exploit complex, high-dimensional cues to excel in-domain, but those cues do not survive cross-domain shifts, so accuracy collapses while ranking may persist.

% ——————————————————————————————————————————————————————————————————————————————————————————
\subsubsection{Distributional Pattern Fingerprints}
\label{subsubsec:dist-fingerprints}
\begin{figure}[t]
  \centering
  \includegraphics[width=\columnwidth]{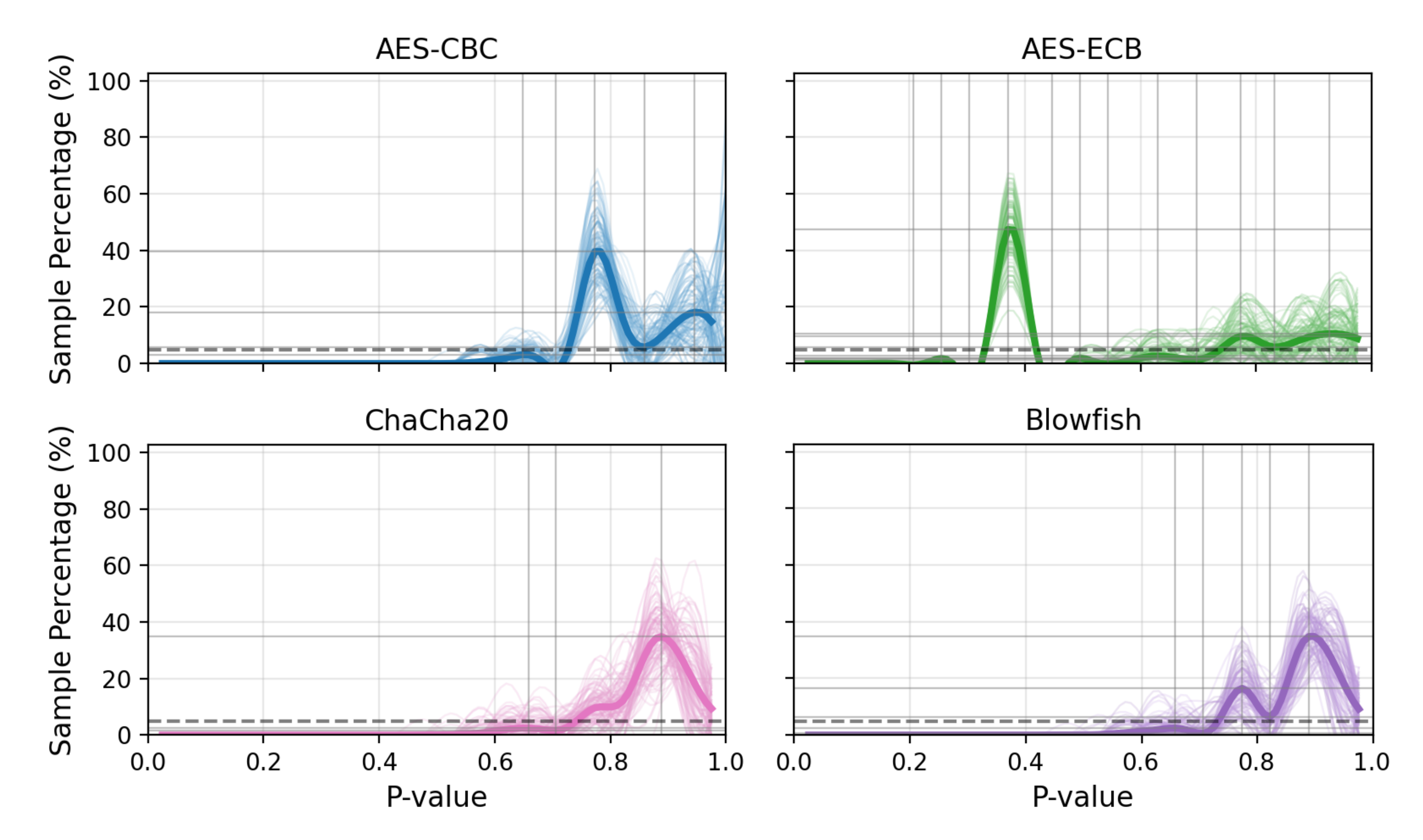}
  \caption{Distributional fingerprints for a \emph{representative subset} of ciphers:
  AES--ECB, AES--CBC, Blowfish, and ChaCha20 (NIST-STS \textit{Non-overlapping} feature).
  Thin lines are 100 seeds; the thick line is the across-seed mean.
  The consistent, cipher-specific shapes underline stability across seeds and
  highlight differences across modes (ECB vs.\ CBC), structures (SPN vs.\ Feistel),
  and cipher families (block vs.\ stream).}
  \label{fig:dist-fingerprint}
\end{figure}

Figure~\ref{fig:dist-fingerprint} shows p-value fingerprints for the NIST-STS \textit{Non-overlapping Template Matching} test across five ciphers: thin lines denote 100 seeds, the thick line is their mean, and the dashed line marks the ideal uniform baseline. Moreover, the curve shapes are cipher-specific. Peaks, valleys, and tails differ by algorithm, and they remain consistent across seeds (and reasonable window sizes), indicating they are not sampling artifacts. Consequently, this consistency helps explain the small cross-domain gaps among the \textit{Regular} datasets in Fig.~\ref{fig:svm-ddr-gap}.
Taken together, these fingerprints offer a model-agnostic explanation: they clarify why our features transfer well across the \textit{Regular} datasets yet face intrinsic limits under the Random\_100 regime.

% ——————————————————————————————————————————————————————————————————————————————————————————
\subsection{Performance Analysis}
In this section, we evaluate the feature extractor on five datasets and nine classifiers to assess both effectiveness and efficiency. The classifiers span linear models (SVM), tree ensembles (Random Forest, XGBoost), logistic regression, and light deep networks (1D ResNet, Transformer).

Across these benchmarks, the results exhibit a clear two-regime pattern (Table~\ref{tab:c_r_m_v}). 
(1) On structured datasets (\texttt{Regular\_100/Regular\_75/Regular\_50}), classical models—linear/RBF SVM, Random Forest, Logistic Regression, and XGBoost—are near saturated. Deep models are competitive on these sets, with ResNet and Transformer matching classical models, but their advantage over simpler learners is negligible. (2) As structure weakens (\texttt{Regular\_25}) and especially on the most random set (\texttt{Random\_100}), performance begins to separate by model family. Classical models maintain high AUC while accuracy and macro F1 fall, whereas the 1D CNN degrades the most.

% ——————————————————————————————————————————————————————————————————————————————————————————
\subsubsection{Performance on structured benchmarks}
Across the structured datasets (\texttt{Regular\_100}, \texttt{Regular\_75}, \texttt{Regular\_50}), the five classical models (linear/RBF SVM, Random Forest, Logistic Regression, XGBoost) achieve saturated performance on all metrics. This suggests that, with our features, the decision boundary is well aligned with class structure and can be captured by standard learners. It further implies that more complex backbones offer limited additional gains over strong classical baselines in this regime.

% ——————————————————————————————————————————————————————————————————————————————————————————
\subsubsection{Performance under reduced structure}
On \texttt{Regular\_25}, XGBoost achieves the best scores, MLP ranks second, and SVMs show only a mild AUC reduction compared to the structured regimes. By contrast, 1D CNN drops markedly, and 1D ResNet also recedes on \texttt{Regular\_25}. These results point to a practical guideline: when local temporal structure weakens, models that leverage global tabular statistics (e.g., tree ensembles, MLP) tend to outperform purely convolutional sequence learners.

% ——————————————————————————————————————————————————————————————————————————————————————————
\subsubsection{Performance on the randomized domain}
On \texttt{Random\_100}, classical models retain high AUC while Acc/F1 move to the 0.52–0.57 range. 1D CNN collapses, whereas 1D ResNet improves ranking yet still struggles on classification. Transformer matches the strongest AUC among learned models, while MLP attains the highest macro F1 by a narrow margin but with a lower AUC. In general, \texttt{Random\_100} exhibits a consistent pattern: ranking quality persists, whereas threshold-based metrics (Accuracy and macro F1) decline due to weaker separability.

Overall, the proposed feature extractor performs strongly across model families and data regimes. On the structured benchmarks (\texttt{Regular\_100/Regular\_75/Regular\_50}), the extracted features are highly discriminative: SVM, Random Forest, Logistic Regression, and XGBoost reach near-saturated Accuracy/F1/AUC, leaving limited headroom for more complex architectures.

On \texttt{Random\_100}, a clear divergence appears: AUC remains high while Accuracy and F1 decrease, indicating that ranking ability is preserved but decision boundaries become less separable under random plaintext. Models that operate on global tabular statistics (SVM, XGBoost) and attention (Transformer) keep AUC around 0.90 to 0.91, whereas convolutional sequence models degrade markedly. Considering accuracy, robustness, and training cost, XGBoost offers the most balanced default for deployment.

\begin{table*}[ht]
\centering
\caption{Performance Evaluation of Classical and Deep Learning Classifiers across Datasets with Varying Plaintext Structure}
\label{tab:c_r_m_v}
\resizebox{\textwidth}{!}{
% --- 修改点 1: 移除了所有 '|' ---
\begin{tabular}{l c ccccc c} 
\toprule
% --- 修改点 2: 移除了 \multicolumn 中的 '|' ---
\multirow{2}{*}{\textbf{Model}} & \multicolumn{1}{c}{\multirow{2}{*}{\textbf{Metric}}} & \multicolumn{5}{c}{\textbf{Datasets}} & \multicolumn{1}{c}{\multirow{2}{*}{\textbf{Overall}}} \\
\cmidrule(lr){3-7}
& & Regular\_100 & Regular\_75 & Regular\_50 & Regular\_25 & Random\_100 & \\
\midrule

\multicolumn{8}{l}{\textbf{Classical learners}} \\
\midrule
\multirow{3}{*}{SVM (linear)}
 & Acc & 0.999 & 0.999 & 0.999 & 0.899 & \bb{0.566} & 0.893 \\
 & F1  & 0.999 & 0.999 & 0.999 & 0.900 & 0.565 & 0.893 \\
 & AUC & 0.999 & 0.999 & 0.999 & 0.988 & \bb{0.915} & \bb{0.981} \\
\midrule
\multirow{3}{*}{SVM (RBF)}
 & Acc & 0.999 & 0.999 & 0.999 & 0.910 & 0.527 & 0.887 \\
 & F1  & 0.999 & 0.999 & 0.999 & 0.912 & 0.524 & 0.887 \\
 & AUC & 0.999 & 0.999 & 0.999 & 0.989 & 0.901 & 0.978 \\
\midrule
\multirow{3}{*}{Logistic Regression}
 & Acc & 0.999 & 0.999 & 0.999 & 0.905 & 0.527 & 0.886 \\
 & F1  & 0.999 & 0.999 & 0.999 & 0.905 & 0.527 & 0.886 \\
 & AUC & 0.999 & 0.999 & 0.999 & 0.991 & 0.817 & 0.962 \\
\midrule
\multirow{3}{*}{Random Forest}
 & Acc & 0.999 & 0.999 & 0.999 & 0.955 & 0.522 & 0.895 \\
 & F1  & 0.999 & 0.999 & 0.999 & 0.955 & 0.512 & 0.893 \\
 & AUC & 0.999 & 0.999 & 0.999 & 0.997 & 0.907 & \bb{0.981} \\
\midrule
\multirow{3}{*}{XGBoost}
 & Acc & 0.999 & 0.999 & 0.999 & 0.983 & 0.541 & \bb{0.905} \\
 & F1  & 0.999 & 0.999 & 0.999 & 0.983 & 0.540 & \bb{0.905} \\
 & AUC & 0.999 & 0.999 & 0.999 & 0.999 & 0.905 & \bb{0.981} \\
\midrule

\multicolumn{8}{l}{\textbf{Deep neural networks}} \\
\midrule
\multirow{3}{*}{1D CNN}
 & Acc & 0.999 & 0.878 & 0.652 & 0.449 & \rr{0.229} & \rr{0.642} \\
 & F1  & 0.999 & 0.879 & 0.643 & 0.438 & \rr{0.177} & \rr{0.627} \\
 & AUC & 0.999 & 0.990 & 0.885 & 0.776 & \rr{0.592} & \rr{0.848} \\
\midrule
\multirow{3}{*}{1D ResNet}
 & Acc & 0.999 & 0.999 & 0.995 & 0.888 & 0.423 & 0.861 \\
 & F1  & 0.999 & 0.999 & 0.995 & 0.888 & 0.366 & 0.850 \\
 & AUC & 0.999 & 0.999 & 0.999 & 0.983 & 0.825 & 0.962 \\
\midrule
\multirow{3}{*}{Transformer}
 & Acc & 0.999 & 0.999 & 0.981 & 0.878 & 0.555 & 0.883 \\
 & F1  & 0.999 & 0.999 & 0.981 & 0.877 & 0.519 & 0.875 \\
 & AUC & 0.999 & 0.999 & 0.998 & 0.977 & 0.914 & 0.978 \\
\midrule
\multirow{3}{*}{MLP}
 & Acc & 0.999 & 0.999 & 0.999 & 0.943 & 0.565 & 0.902 \\
 & F1  & 0.999 & 0.999 & 0.999 & 0.942 & \bb{0.566} & 0.902 \\
 & AUC & 0.999 & 0.999 & 0.999 & 0.994 & 0.844 & 0.968 \\
\bottomrule
\end{tabular}}
\end{table*}
% ——————————————————————————————————————————————————————————————————————————————————————————————
\subsection{Comparative Evaluation}

To assess the effectiveness of our feature extractor, we compare it with several strong baseline feature sets. The baselines include RawP (a raw feature baseline), KS inspired (a statistical baseline), and their union RawP+KS. For a fair and comprehensive comparison, each feature set is paired with three representative learners (SVM, XGBoost, and MLP) and evaluated on all five datasets. We use macro AUC and macro F1 as the primary metrics.

\begin{figure}[t]
  \centering
  \includegraphics[width=\columnwidth]{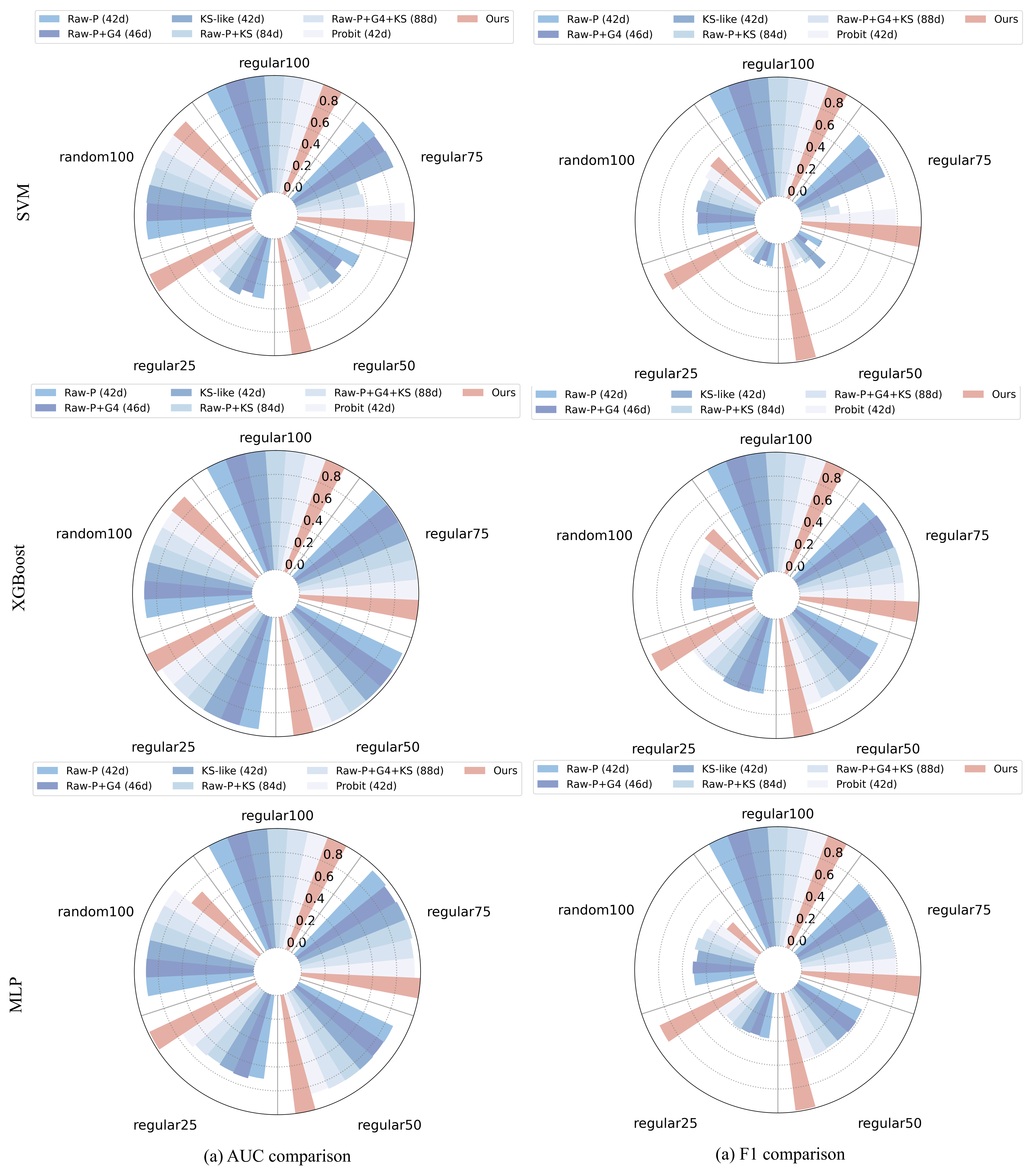}
  \caption{Comparative results with our feature extractor. Left: macro AUC; right: macro F1. SVM/XGBoost consistently lead across datasets, while MLP is competitive in-domain but degrades on \emph{\texttt{Random\_100}}.}
  \label{fig:compare}
\end{figure}

Figure~\ref{fig:compare} summarizes the results with radar charts; across datasets and metrics, our feature extractor (Ours) consistently outperforms the baseline feature sets. In the radar plots, the red polygon for Ours spans the largest area, indicating better performance on most datasets and metrics. This pattern holds across the three learners. Across SVM, XGBoost, and MLP, Ours yields the top scores among the compared feature sets. The largest margin appears on the \texttt{Random\_100} dataset. On this challenging cross-domain benchmark, RawP and KS-inspired drop sharply, with macro F1 approaching chance levels. By contrast, pairing Ours with SVM or XGBoost keeps macro AUC above 0.90, indicating stronger tolerance to domain shift. MLP remains competitive in-domain but drops more on \texttt{Random\_100}, especially in macro F1. Across SVM, XGBoost, and MLP, the relative ranking of feature sets is consistent, which suggests the gains come from the representation rather than any single model. We next discuss interpretation and implications.

The comparative evaluation reveals a decisive and universal superiority of our proposed feature extractor. This advantage is most pronounced under the significant domain shift presented by the \texttt{Random\_100} dataset, where baseline feature sets collapse to near-chance performance. In contrast, our features maintain high ranking ability (macro AUC $>$ 0.90) across multiple classifiers. The consistent performance lift observed with diverse models—from SVM and XGBoost to MLP—provides compelling evidence that the improvement stems from the fundamental power of the representation itself, not a model-specific artifact.

These results highlight a critical limitation in conventional feature design for this task. Baselines like RawP are susceptible to learning spurious correlations tied to the training data's specific plaintext structure, which fail to generalize. Similarly, statistical features like KS inspired, while useful, often rely on distributional moments that are not invariant to content shifts. Our approach overcomes these weaknesses by capturing more fundamental and stable properties of the cryptographic transformation. This finding advocates for a methodological shift in the field: rather than relying on increasingly complex models to compensate for brittle features, the focus should be on engineering inherently robust representations first. For practitioners, this translates to a more reliable and model-agnostic foundation, simplifying the development of effective, real-world identification systems.

% ——————————————————————————————————————————————————————————————————————————————————————————
\subsection{Ablation Study}
In this section, we perform ablation studies to verify the effectiveness of important modules. We conduct a comprehensive ablation study to further determine the efficacy of each key component. Specifically, we compare the full method with three variants:
\begin{itemize}
  \item \textbf{Statistical features only:} Uses only the $S$ summary scalars per test (mean, variance, skewness, kurtosis); the histogram channel is removed.
  \item \textbf{Histogram features only:} Uses only the $K$ bin normalized histogram per test; the summary scalars are removed.
  \item \textbf{Reduced order moments:} Keeps mean and variance while removing skewness and kurtosis; the histogram channel is retained.
\end{itemize}

\begin{figure*}[!t]
  \centering
  \includegraphics[width=\textwidth, trim=2pt 2pt 2pt 2pt, clip]{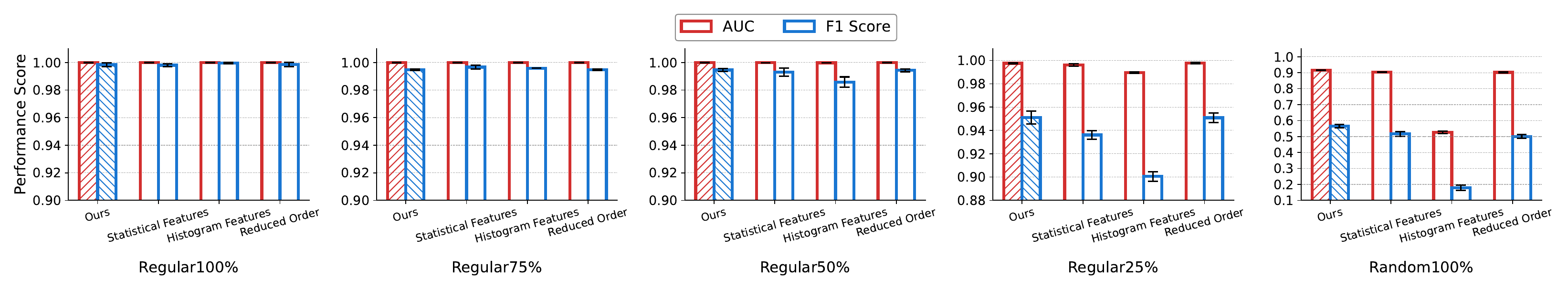}
  \caption{Ablation across datasets. ``Ours'' uses all features; ``Statistical features'' uses mean, variance, skewness, and kurtosis; ``Histogram features'' use the $K$-bin normalized histogram; ``Reduced order'' removes the higher-order moments (skewness and kurtosis).}
  \label{fig:ablation}
\end{figure*}

Figure~\ref{fig:ablation} shows that on structured datasets (\texttt{Regular\_100}, \texttt{Regular\_75}, \texttt{Regular\_50}) all variants perform well, with the full model holding a small but consistent advantage. At \texttt{Regular\_25}, the gaps widen; on the most challenging \texttt{Random\_100} set, single component variants drop sharply while the full model sustains high AUC and F1. These trends suggest that as regularity decreases, any single feature family becomes insufficient.

Why does the full model remain strong under stress? Each feature family targets a different part of the signal: the summary scalars capture global trends, the histogram encodes local density, and the higher order moments emphasize tail behavior. On their own these views are fragile under domain shift, whereas together they form a more complete and more stable representation.

Each component adds value, and their combination yields the strongest robustness. For deployment, we recommend the full feature set. When latency or memory is a constraint on structured workloads, removing the higher order moments (skewness and kurtosis) has little effect, but we do not recommend this simplification when cross-domain generalization is required.

% ——————————————————————————————————————————————————————————————————————————————————————————
\subsection{Robustness Evaluation}

\begin{figure}[t]
  \centering
  \captionsetup{font=small}
  \includegraphics[width=\columnwidth]{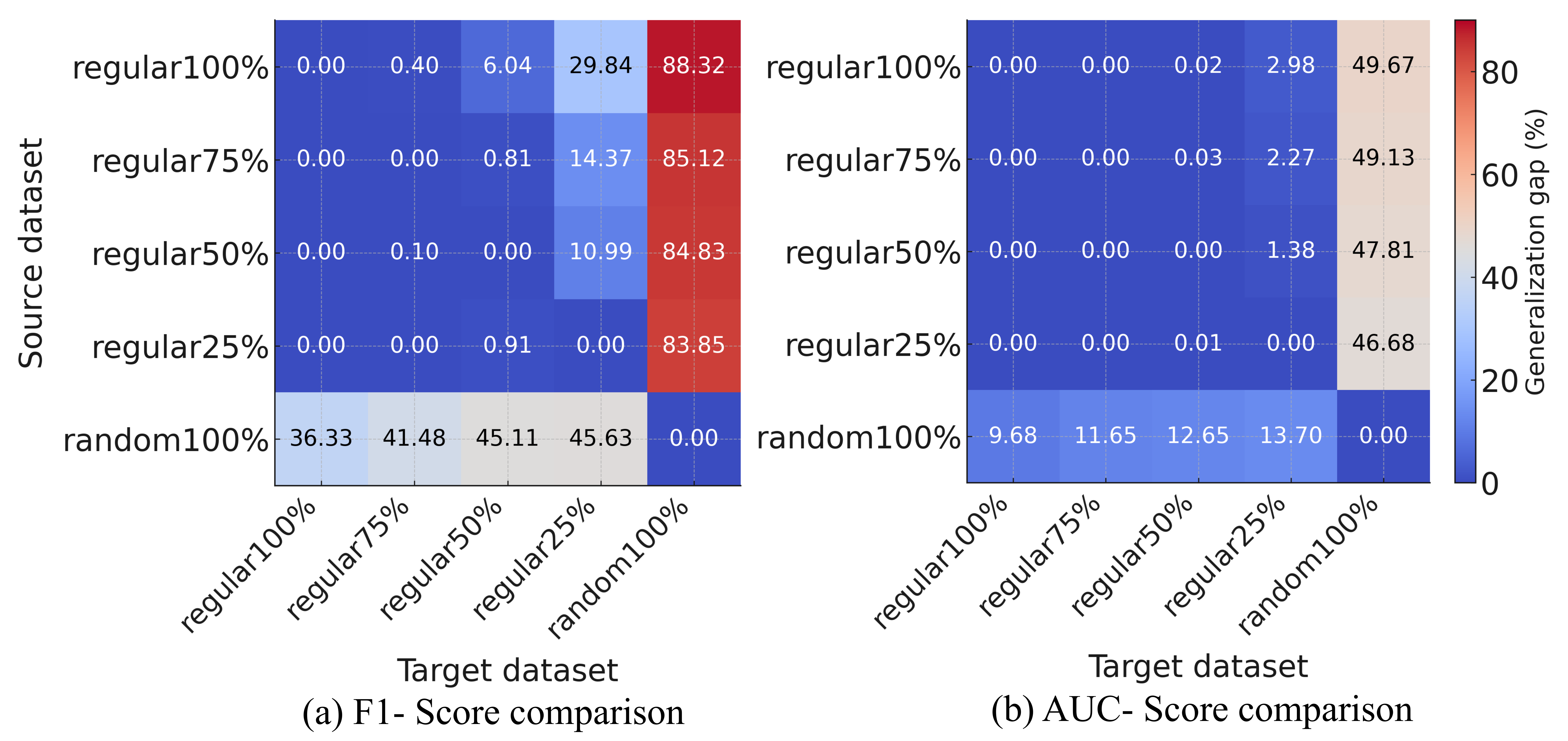}
  \caption{Cross-domain generalization gaps (averaged across all classifiers).
  Each cell shows the percentage drop relative to in-domain performance (0\% on the diagonal).
  Warmer colors indicate larger gaps. Transfers into \texttt{Random\_100} exhibit the largest gaps,
  whereas transfers within the Regular regime remain small.}
  \label{fig:svm-ddr-gap}
\end{figure}

% --- 修改后的跨双栏图表 ---

\begin{figure*}[t]  % <-- 1. 从 figure 修改为 figure*
  \centering
  \captionsetup{font=small}
  \includegraphics[width=\textwidth]{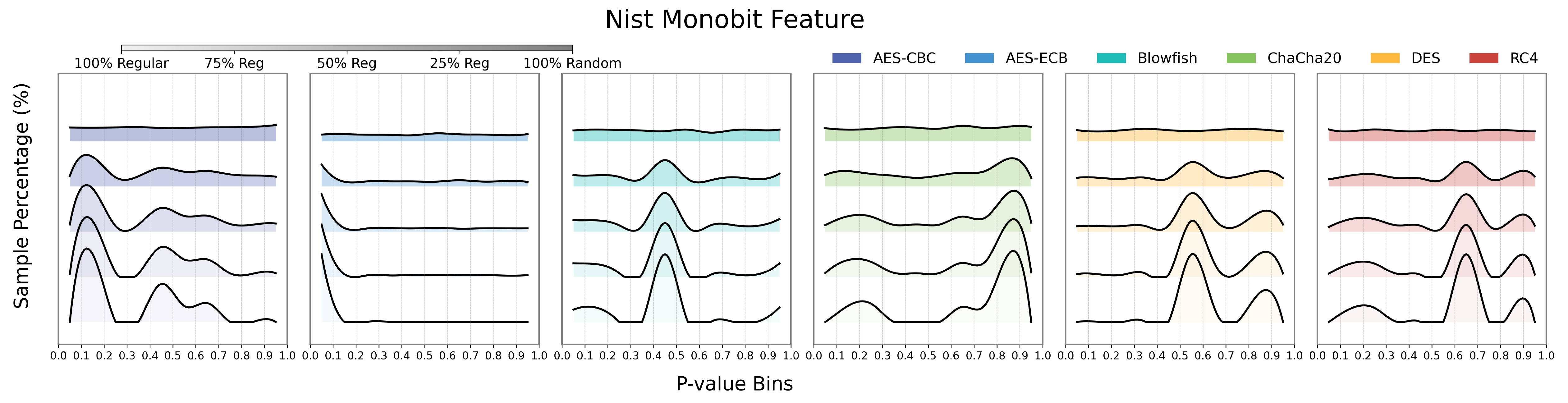} % <-- 2. 从 \columnwidth 修改为 \textwidth
  \caption{Distributional fingerprints for a representative NIST-STS feature (Monobit).
  Ridgelines show the density over p-value bins from \texttt{Regular\_100} to \texttt{Random\_100}.
  Curves are stable within Regular datasets but flatten and overlap toward \texttt{Random\_100}, aligning with the larger average gaps.}
  \label{fig:dist-fp-monobit}
\end{figure*} % <-- 3. 对应修改为 \end{figure*}

In this section we evaluate robustness under domain shift using the cross-domain generalization gap (Eq.~\ref{eq:gap}), training on a source dataset and testing on a different target without any fine-tuning (T0).
Figure~\ref{fig:svm-ddr-gap} reports average gaps across all classifiers; each cell is the percentage drop relative to in-domain performance (0\% on the diagonal).
Within the Regular regime (\texttt{Regular\_100}/\texttt{Regular\_75}/\texttt{Regular\_50}/\texttt{Regular\_25}), gaps remain small: macro-F1 decreases by roughly 0.00--6.04\% and macro-AUC by 0.00--2.98\% on average.

By contrast, transfers into the \texttt{Random\_100} target yield much larger average gaps: macro-F1 drops by about 83.85--88.32\%, whereas macro-AUC declines by 46.68--49.67\%.
Although both metrics degrade substantially, the F1 drop is consistently larger than the AUC drop, suggesting that some ranking ability remains even when class boundaries blur.
To interpret these gaps, we examine distributional fingerprints of calibrated test scores.

Figure~\ref{fig:dist-fp-monobit} shows ridgeline plots for the NIST-STS Monobit feature from \texttt{Regular\_100} to \texttt{Random\_100}.
Within the Regular regime, curves are cipher-specific yet stable across seeds, with clear peaks and valleys; the shapes change only mildly across Regular datasets, consistent with the small gaps.
Toward \texttt{Random\_100}, the curves flatten and increasingly overlap, reducing separability and aligning with the large average gaps in Fig.~\ref{fig:svm-ddr-gap}.

Overall, the averaged results indicate strong robustness under moderate shifts within the Regular regime and pronounced sensitivity when the target is \texttt{Random\_100}.
Moreover, the distributional fingerprints provide a model-agnostic explanation for these trends, linking shape stability to small gaps and shape flattening to large gaps.

% ——————————————————————————————————————————————————————————————————————————————————————————
% ——————————————————————————————————————————————————————————————————————————————————————————
\subsection{Benchmark Evaluation on the Public Canterbury Dataset}
\label{sec:canterbury}

To validate our features against standard recipes on a real-world public benchmark, we first present a comparative evaluation and ablation study on the Canterbury corpus. We use three representative classifiers (SVM, XGBoost, and MLP) to ensure the findings are not tied to a specific model. Key results are summarized in Table~\ref{tab:canterbury_comp_ablation}. The left block in Table~\ref{tab:canterbury_comp_ablation} benchmarks six widely used recipes. The right block conducts ablations on our feature design: \textit{Only\_Bins} (histogram channel only), \textit{Only\_Stats} (summary scalars only), and \textit{No\_HighOrder} (drop skewness/kurtosis). ``Ours'' denotes the full proposed feature set.

Table~\ref{tab:canterbury_comp_ablation} shows consistent gains of ``Ours'' over the six recipes across all three models. 
On SVM (RBF), ``Ours'' achieves \textbf{0.897} $\pm$ \textbf{0.071} Acc, \textbf{0.899} $\pm$ \textbf{0.067} F1, and \textbf{0.985} $\pm$ \textbf{0.025} AUC, clearly outperforming the best recipe column. The trend holds for XGBoost (\textbf{0.788}/\textbf{0.782}/\textbf{0.952}) and MLP (\textbf{0.878}/\textbf{0.876}/\textbf{0.982}), indicating that our features transfer across model families.

Across all models, \textit{Only\_Bins} ranks second and is much stronger than \textit{Only\_Stats}, 
showing that the normalized histogram channel captures discriminative local density patterns beyond global moments. 
Removing higher-order moments (\textit{No\_HighOrder}) further degrades performance compared with ``Ours'', 
confirming that skewness and kurtosis provide complementary tail sensitivity. 
In short, summary scalars model global trends; histograms encode local density; higher-order moments emphasize tail behavior.
Their combination yields the most robust representation.

(i) Our proposed features are competitive on a public corpus with grouped CV, not only on our in-house datasets. 
(ii) Histogram features are the main driver, but higher-order moments consistently add value; 
dropping them is only recommended when latency or memory is critical and domain shift is limited. 
(iii) The relative ordering of recipes is stable across SVM/XGBoost/MLP, suggesting that the gains stem from features rather than model choice.

% ——————————————————————————————————————Canterbury————————————————————————————————————————————————
\begin{table*}[t]
\caption{Overall Results on the Canterbury Dataset.
Shorthands: \textbf{G4} = Global-4 moments (mean, variance, skewness, excess kurtosis) computed per test column; 
\textbf{KS} = per-test one-sample KS statistic vs.\ $\mathrm{Uniform}(0,1)$.}
\label{tab:canterbury_comp_ablation}
\centering
\scriptsize
\setlength{\tabcolsep}{2.5pt}
\renewcommand{\arraystretch}{1.08}

\resizebox{\textwidth}{!}{%
\begin{tabular}{ll*{6}{c}*{3}{c}c}
\toprule
\multirow{2}{*}{\textbf{Model}} & \multirow{2}{*}{\textbf{Metric}} &
\multicolumn{6}{c}{\textbf{Comparison}} &
\multicolumn{3}{c}{\textbf{Ablation}} &
\textbf{Ours} \\
\cmidrule(lr){3-8}\cmidrule(lr){9-11}\cmidrule(lr){12-12}
& & Raw\_P & KS & Raw\_P+G4 & Raw\_P+KS & Raw\_P+G4+KS & Probit
  & Only\_Bins & Only\_Stats & No\_HighOrder & Baseline \\
\midrule
\multirow{3}{*}{SVM}
& Acc & $0.219\pm0.063$ & $0.206\pm0.054$ & $0.214\pm0.064$ & $0.218\pm0.063$ & $0.221\pm0.054$ & $0.221\pm0.052$ & $0.879\pm0.022$ & $0.772\pm0.017$ & $0.727\pm0.016$ & $\textbf{0.897}\pm\textbf{0.071}$ \\
& F1  & $0.204\pm0.058$ & $0.186\pm0.047$ & $0.198\pm0.058$ & $0.199\pm0.057$ & $0.200\pm0.050$ & $0.208\pm0.049$ & $0.877\pm0.022$ & $0.768\pm0.019$ & $0.721\pm0.012$ & $\textbf{0.899}\pm\textbf{0.067}$ \\
& AUC & $0.577\pm0.040$ & $0.570\pm0.061$ & $0.579\pm0.035$ & $0.577\pm0.056$ & $0.577\pm0.055$ & $0.603\pm0.030$ & $0.982\pm0.007$ & $0.948\pm0.005$ & $0.928\pm0.005$ & $\textbf{0.985}\pm\textbf{0.025}$ \\
\midrule
\multirow{3}{*}{XGBoost}
& Acc & $0.211\pm0.066$ & $0.228\pm0.033$ & $0.222\pm0.054$ & $0.221\pm0.044$ & $0.232\pm0.041$ & $0.205\pm0.051$ & $0.770\pm0.030$ & $0.614\pm0.021$ & $0.576\pm0.032$ & $\textbf{0.788}\pm\textbf{0.059}$ \\
& F1  & $0.207\pm0.055$ & $0.221\pm0.019$ & $0.217\pm0.038$ & $0.213\pm0.030$ & $0.224\pm0.030$ & $0.198\pm0.040$ & $0.763\pm0.031$ & $0.608\pm0.016$ & $0.589\pm0.033$ & $\textbf{0.782}\pm\textbf{0.058}$ \\
& AUC & $0.570\pm0.037$ & $0.585\pm0.022$ & $0.571\pm0.034$ & $0.573\pm0.029$ & $0.577\pm0.027$ & $0.559\pm0.039$ & $0.944\pm0.007$ & $0.886\pm0.007$ & $0.879\pm0.009$ & $\textbf{0.952}\pm\textbf{0.025}$ \\
\midrule
\multirow{3}{*}{MLP}
& Acc & $0.227\pm0.061$ & $0.213\pm0.059$ & $0.195\pm0.050$ & $0.205\pm0.067$ & $0.204\pm0.053$ & $0.230\pm0.060$ & $0.837\pm0.067$ & $0.716\pm0.012$ & $0.757\pm0.011$ & $\textbf{0.878}\pm\textbf{0.030}$ \\
& F1  & $0.218\pm0.060$ & $0.201\pm0.049$ & $0.185\pm0.056$ & $0.199\pm0.069$ & $0.190\pm0.052$ & $0.220\pm0.062$ & $0.836\pm0.064$ & $0.713\pm0.011$ & $0.756\pm0.013$ & $\textbf{0.876}\pm\textbf{0.031}$ \\
& AUC & $0.581\pm0.054$ & $0.564\pm0.064$ & $0.570\pm0.036$ & $0.554\pm0.054$ & $0.571\pm0.061$ & $0.590\pm0.042$ & $0.961\pm0.032$ & $0.925\pm0.006$ & $0.938\pm0.008$ & $\textbf{0.982}\pm\textbf{0.007}$ \\
\bottomrule
\end{tabular}%
}
\end{table*}
% ——————————————————————————————————————Canterbury————————————————————————————————————————————————
% ——————————————————————————————————————————————————————————————————————————————————————————

\section{DISCUSSION}
\label{sec:discussion}

\subsection{Principal Findings}
The foremost finding of this study is the Plaintext Structure Vulnerability, a core weakness in current cryptographic algorithm identification. Our experiments revealed that models trained on structured plaintext degrade sharply when evaluated on random plaintext. This indicates that many state of the art models learn content statistics rather than intrinsic algorithmic signals. Instead, they exploit statistical artifacts from the plaintext, which limits reliability in real-world applications. We also observe that simpler models, such as XGBoost, can be more robust than deeper architectures: greater capacity may increase sensitivity to content specific patterns.
% ——————————————————————————————————————————————————————————————————————————————————————————
\subsection{Performance and SNR}
\label{subsec:snr_interpretation}
Our key findings on the \texttt{Random\_100} dataset can be explained from the perspective of signal-to-noise ratio (SNR). From this perspective, the statistical fingerprint of the algorithm is a signal, while the structure of the plaintext is noise, so the structured plaintext provides a low noise background (high SNR), so that the feature distribution has good separability, thus bringing high accuracy. In contrast, the random text produces a high entropy background of an approximate signal, thus reducing the SNR. According to the detection theory, a lower SNR will increase the overlap between the class conditional distributions, which sets an upper limit for performance (i.e. Bayesian error rate), and the upper limit has nothing to do with the classifier used~\cite{kay1998detection, bishop2006prml, cover1999elements}. This perspective also explains the difference between metrics: while threshold based metrics such as accuracy and F1 score drop with increased overlap, ranking based AUC can remain relatively high because a weak but persistent signal still supports better than chance ranking~\cite{fawcett2006roc}. Therefore, the limit observed on \texttt{Random\_100} reflects an intrinsic information constraint rather than a failure of the feature design.
% ——————————————————————————————————————————————————————————————————————————————————————————
\subsection{Implications for the Field}
These results suggest that evaluation should not focus solely on accuracy within homogeneous datasets. Robustness to variations in plaintext structure should be a primary criterion for security applications. For practitioners, a classifier that reports high accuracy in a controlled setting may underperform in deployment where plaintext characteristics vary widely. Incorporating robustness checks across plaintext regimes can reduce this risk.
% ——————————————————————————————————————————————————————————————————————————————————————————
\subsection{Limitations}
This study has several limitations. First, experiments cover six common cryptographic algorithms; future work should include a broader set of ciphers, including less common and proprietary variants. Second, the composition of the statistical test suite and the extractor hyperparameters (e.g., window size $W$, number of histogram bins $K$) were fixed; sensitivity to these choices merits further study. Finally, the datasets are synthetic and designed to isolate plaintext structure effects; validation on captured network traffic is needed to account for operational complexities.
% ——————————————————————————————————————————————————————————————————————————————————————————
\subsection{Future Work}
Future directions include expanding the cipher library and testing on real-world, large scale traffic. Another direction is to select or learn an optimal test suite automatically rather than using a fixed set. In addition, adapting the window based feature extraction to online and streaming scenarios would support practical deployment. More broadly, distributional fingerprints may benefit other security tasks beyond cryptographic identification.
% ——————————————————————————————————————————————————————————————————————————————————————————
\section{CONCLUSION}
\label{sec:conclusion}
In this paper, we identify plaintext structure vulnerability as a critical flaw in cryptographic identification that causes models to analyze using misleading plaintext features rather than inherent cryptographic features. To address this problem, we propose a robust distributional randomness fingerprinting that lifts the analysis from the volatile raw bytes to a more stable statistical domain. Experiments confirm that this feature engineering first approach keeps high performance in serious domain shift scenarios where the byte-level model fails. The core insight is that the field must adopt “robustness to plaintext variation” as a required evaluation metric, breaking through the limitations of homogenized benchmarking. This approach not only provides a more reliable identification tool, but also opens the way to apply similar statistical fingerprinting to a wider range of security challenges, such as weak random number generator detection or cryptographic traffic analysis.

\section*{Acknowledgment}

This work was supported by the National Cryptologic Science Fund of China
(Grant No. 2025NCSF02019).

\bibliographystyle{IEEEtran}
\bibliography{IEEEabrv,bib}

\begin{thebibliography}{10}
\providecommand{\url}[1]{#1}
\csname url@samestyle\endcsname
\providecommand{\newblock}{\relax}
\providecommand{\bibinfo}[2]{#2}
\providecommand{\BIBentrySTDinterwordspacing}{\spaceskip=0pt\relax}
\providecommand{\BIBentryALTinterwordstretchfactor}{4}
\providecommand{\BIBentryALTinterwordspacing}{\spaceskip=\fontdimen2\font plus
\BIBentryALTinterwordstretchfactor\fontdimen3\font minus
  \fontdimen4\font\relax}
\providecommand{\BIBforeignlanguage}[2]{{%
\expandafter\ifx\csname l@#1\endcsname\relax
\typeout{** WARNING: IEEEtran.bst: No hyphenation pattern has been}%
\typeout{** loaded for the language `#1'. Using the pattern for}%
\typeout{** the default language instead.}%
\else
\language=\csname l@#1\endcsname
\fi
#2}}
\providecommand{\BIBdecl}{\relax}
\BIBdecl

\bibitem{meijer2021s}
C.~Meijer, V.~Moonsamy, and J.~Wetzels, ``Where's crypto?: Automated
  identification and classification of proprietary cryptographic primitives in
  binary code,'' in \emph{30th USENIX Security Symposium (USENIX Security 21)},
  2021, pp. 555--572.

\bibitem{nist2019fips3}
\BIBentryALTinterwordspacing
{National Institute of Standards and Technology}, ``Fips {PUB} 140-3: Security
  requirements for cryptographic modules,'' U.S. Department of Commerce, Tech.
  Rep. FIPS PUB 140-3, March 2019, online; available at URL. [Online].
  Available: \url{https://csrc.nist.gov/publications/detail/fips/140/3/final}
\BIBentrySTDinterwordspacing

\bibitem{erwradar2025}
L.~Zhao, Y.~Zhang, Z.~Wang, F.~Yuan, and R.~Hou, ``Erw-radar: An adaptive
  detection system against evasive ransomware by contextual behavior detection
  and fine-grained content analysis.'' in \emph{NDSS}, 2025.

\bibitem{wang2024cryptody}
J.~Wang, S.~Guo, W.~Diao, Y.~Liu, H.~Duan, Y.~Liu, and Z.~Liang, ``Cryptody:
  Cryptographic misuse analysis of iot firmware via data-flow reasoning,'' in
  \emph{Proceedings of the 27th International Symposium on Research in Attacks,
  Intrusions and Defenses}, 2024, pp. 579--593.

\bibitem{mousavi2025detecting}
Z.~Mousavi, C.~Islam, M.~A. Babar, A.~Abuadbba, and K.~Moore, ``Detecting
  misuse of security apis: A systematic review,'' \emph{ACM Computing Surveys},
  vol.~57, no.~12, pp. 1--39, 2025.

\bibitem{ami2022crypto}
A.~S. Ami, N.~Cooper, K.~Kafle, K.~Moran, D.~Poshyvanyk, and A.~Nadkarni, ``Why
  crypto-detectors fail: A systematic evaluation of cryptographic misuse
  detection techniques,'' in \emph{2022 IEEE Symposium on Security and Privacy
  (SP)}.\hskip 1em plus 0.5em minus 0.4em\relax IEEE, 2022, pp. 614--631.

\bibitem{bincrypto2025}
Y.~Hu, Y.~He, W.~He, H.~Li, Y.~Zhao, S.~Wang, and D.~Gu, ``Binary cryptographic
  function identification via similarity analysis with path-insensitive
  emulation,'' \emph{Proceedings of the ACM on Programming Languages}, vol.~9,
  no. OOPSLA1, pp. 28--56, 2025.

\bibitem{foc2024}
X.~Shang, G.~Chen, S.~Cheng, S.~Guo, Y.~Zhang, W.~Zhang, and N.~Yu, ``Foc:
  Figure out the cryptographic functions in stripped binaries with llms,''
  \emph{ACM Transactions on Software Engineering and Methodology}, 2024.

\bibitem{li2025generic}
J.~Li, H.~Sun, Z.~Du, Y.~Wang, K.~Yuan, and C.~Jia, ``A generic cryptographic
  algorithm identification scheme based on ciphertext features,'' \emph{Journal
  of Information Security and Applications}, vol.~89, p. 103984, 2025.

\bibitem{hu2025cnn}
H.~Hu and K.~Yuan, ``Identification of cryptographic algorithms based on cnn,''
  in \emph{Proceedings of the 4th International Conference on Computer,
  Artificial Intelligence and Control Engineering}, 2025, pp. 182--186.

\bibitem{yuan2023}
K.~Yuan, Y.~Huang, J.~Li, C.~Jia, and D.~Yu, ``A block cipher algorithm
  identification scheme based on hybrid random forest and logistic regression
  model,'' \emph{Neural Processing Letters}, vol.~55, no.~3, pp. 3185--3203,
  2023.

\bibitem{yuan2025mlp}
K.~Yuan, D.~Yu, W.~Yang, Z.~Du, L.~Shen, and Z.~Li, ``Identification of block
  cipher algorithms using multi-layer perception algorithm,'' \emph{Soft
  Computing}, pp. 1--12, 2025.

\bibitem{dileep2006svm}
A.~D. Dileep and C.~C. Sekhar, ``Identification of block ciphers using support
  vector machines,'' in \emph{The 2006 IEEE International Joint Conference on
  Neural Network Proceedings}.\hskip 1em plus 0.5em minus 0.4em\relax IEEE,
  2006, pp. 2696--2701.

\bibitem{manjula2011dt}
R.~Manjula and R.~Anitha, ``Identification of encryption algorithm using
  decision tree,'' in \emph{International Conference on Computer Science and
  Information Technology}.\hskip 1em plus 0.5em minus 0.4em\relax Springer,
  2011, pp. 237--246.

\bibitem{xie2024}
R.~Xie, X.~Chen, X.~Zhang, and G.~Shi, ``Block cipher algorithm identification
  based on cnn-transformer fusion model,'' in \emph{Chinese Conference on
  Pattern Recognition and Computer Vision (PRCV)}.\hskip 1em plus 0.5em minus
  0.4em\relax Springer, 2024, pp. 97--110.

\bibitem{dani2024machine}
J.~Dani, K.~Nakka, and N.~Saxena, ``A machine learning-based framework for
  assessing cryptographic indistinguishability of lightweight block ciphers,''
  \emph{arXiv preprint arXiv:2405.19683}, 2024.

\bibitem{li2022genda}
X.~Li, Y.~Chang, G.~Ye, X.~Gong, and Z.~Tang, ``Genda: A graph embedded network
  based detection approach on encryption algorithm of binary program,''
  \emph{Journal of Information Security and Applications}, vol.~65, p. 103088,
  2022.

\bibitem{li2018khunt}
J.~Li, Z.~Lin, J.~Caballero, Y.~Zhang, and D.~Gu, ``K-hunt: Pinpointing
  insecure cryptographic keys from execution traces,'' in \emph{Proceedings of
  the 2018 ACM SIGSAC Conference on Computer and Communications Security},
  2018, pp. 412--425.

\bibitem{rukhin2010niststs}
L.~E. Bassham~III, A.~L. Rukhin, J.~Soto, J.~R. Nechvatal, M.~E. Smid, E.~B.
  Barker, S.~D. Leigh, M.~Levenson, M.~Vangel, D.~L. Banks \emph{et~al.}, ``Sp
  800-22 rev. 1a. a statistical test suite for random and pseudorandom number
  generators for cryptographic applications,'' 2010.

\bibitem{brown2006dieharder}
\BIBentryALTinterwordspacing
R.~G. Brown, D.~Eddelbuettel, and D.~Bauer, ``Dieharder: A random number test
  suite,'' 2011--2025, software; accessed 2025-11-10. [Online]. Available:
  \url{https://webhome.phy.duke.edu/~rgb/General/dieharder.php}
\BIBentrySTDinterwordspacing

\bibitem{lecuyer2007testu01}
P.~L'ecuyer and R.~Simard, ``Testu01: Ac library for empirical testing of
  random number generators,'' \emph{ACM Transactions on Mathematical Software
  (TOMS)}, vol.~33, no.~4, pp. 1--40, 2007.

\bibitem{vuursteen2023optimal}
L.~Vuursteen, B.~Szab{\'o}, A.~van~der Vaart, and H.~Van~Zanten, ``Optimal
  testing using combined test statistics across independent studies,''
  \emph{Advances in Neural Information Processing Systems}, vol.~36, pp.
  80\,661--80\,673, 2023.

\bibitem{liu2020cct}
Y.~Liu and J.~Xie, ``Cauchy combination test: a powerful test with analytic
  p-value calculation under arbitrary dependency structures,'' \emph{Journal of
  the American Statistical Association}, vol. 115, no. 529, pp. 393--402, 2020.

\bibitem{koh2021wilds}
P.~W. Koh, S.~Sagawa, H.~Marklund, S.~M. Xie, M.~Zhang, A.~Balsubramani, W.~Hu,
  M.~Yasunaga, R.~L. Phillips, I.~Gao \emph{et~al.}, ``Wilds: A benchmark of
  in-the-wild distribution shifts,'' in \emph{International conference on
  machine learning}.\hskip 1em plus 0.5em minus 0.4em\relax PMLR, 2021, pp.
  5637--5664.

\bibitem{canterbury}
R.~Arnold and T.~Bell, ``A corpus for the evaluation of lossless compression
  algorithms,'' in \emph{Proceedings DCC'97. Data Compression
  Conference}.\hskip 1em plus 0.5em minus 0.4em\relax IEEE, 1997, pp. 201--210.

\bibitem{rfc5869}
H.~Krawczyk and P.~Eronen, ``Hmac-based extract-and-expand key derivation
  function (hkdf),'' Tech. Rep., 2010.

\bibitem{kay1998detection}
S.~M. Kay, ``Fundamentals of statistical signal processing, volume 2: Detection
  theory. 1998.''

\bibitem{bishop2006prml}
C.~M. Bishop and N.~M. Nasrabadi, \emph{Pattern recognition and machine
  learning}.\hskip 1em plus 0.5em minus 0.4em\relax Springer, 2006, vol.~4,
  no.~4.

\bibitem{cover1999elements}
T.~M. Cover, \emph{Elements of information theory}.\hskip 1em plus 0.5em minus
  0.4em\relax John Wiley \& Sons, 1999.

\bibitem{fawcett2006roc}
T.~Fawcett, ``An introduction to roc analysis,'' \emph{Pattern recognition
  letters}, vol.~27, no.~8, pp. 861--874, 2006.

\end{thebibliography}

\vfill

\end{document}